\documentstyle[12pt]{article}

\newcommand{\newsection}{
\setcounter{equation}{0}
\section}
\newcommand{\tr}[1]{\,{\rm tr}\,#1\,}
\def\e{{\,\rm e}\,}
\def\pint{\int\hspace{-1.07em}\not\hspace{0.6em}}

\def\eop{\vspace*{\fill}\pagebreak}
\def\be{\begin{equation}}
\def\ee{\end{equation}}
\def\bea{\begin{eqnarray}}
\def\eea{\end{eqnarray}}
\def\qqqq#1{\sqrt{#1^2+b#1+c}}
\newcommand{\ppp}[1]{(#1^2+b#1+c)}
\def\L{\Lambda}
\def\l{\lambda}
\def\h{\eta}
\def\hm{the hermitian one-matrix model}

\newcommand{\dd}[1]{{\partial \over \partial #1}}
\newcommand{\ddt}[1]{{\partial \over \partial t_{#1}}}
\newcommand{\hf}{{\textstyle{1\over 2}}}

\newcommand{\half}{{\textstyle{1\over 2}}}
\renewcommand{\d}{{{\partial}}}
\newcommand{\p}{^{\prime}}
\newcommand{\ra}{\rightarrow}
\newcommand{\KK}{{\cal K}}

\newcommand{\fr}[2]{{\textstyle {#1 \over #2}}}

\begin{document}

\begin{flushright}
NBI-HE-92-22 \\ February, 1992
\end{flushright}

\begin{center}
{\huge Higher Genus Correlators from \\
\vspace{.5cm} the Hermitian One-Matrix Model}
\end{center}
\vspace{.5cm}
\begin{center}
{\large J. Ambj{\o}rn}\footnote{E--mail: \ ambjorn@nbivax.nbi.dk}  \\
 \mbox{} \\{\it The Niels Bohr Institute} \\
{\it Blegdamsvej 17, DK-2100 Copenhagen, Denmark} \\
\vspace{.5cm} \mbox{}\\
{\large L. Chekhov}\footnote{E--mail: \ chekhov@qft.mian.su}  \\
 \mbox{} \\ {\it Steklov Mathematical Institute} \\
{\it Vavilov st.42, GSP-1, 117966 Moscow, Russia} \\
\vspace{0.5cm} \mbox{} \\
{\large Yu. Makeenko}\footnote{E--mail: \ makeenko@nbivax.nbi.dk \ \ / \ \
makeenko@desyvax.bitnet \ \ / \ \ makeenko@itep.msk.su} \\ \mbox{} \\
{\it Institute of Theoretical and Experimental Physics} \\
{\it B.Cheremuskinskaya 25, 117259 Moscow, Russia}
\end{center}

\vspace{0.5cm}

\begin{abstract}
We develop an iterative algorithm for the genus expansion
of the hermitian $N\times N$ one-matrix model ($=$ the Penner model in
an external field). By introducing moments of the external field, we prove
that the genus $g$ contribution to the $m$-loop correlator depends only
on $3g-2+m$ lower moments ($3g-2$ for the partition function).
We present the explicit results for the partition function and the one-loop
correlator in genus one. We compare the correlators
for the hermitian one-matrix model with those at zero momenta for $c=1$ CFT
and show an agreement of the one-loop correlators for genus zero.
\end{abstract}

\noindent
Submitted to {\sl Physics Letters B}

\eop

\newsection {Introduction}

Recently there has renewed interest in \hm\ which is defined by the partition
function
\be
Z_N [t.] = \int DX\, \e^{-\sum_{k=0}^\infty t_k \tr{X^k}}
\label{1}
\ee
where the integration goes over $N\times N$ hermitian
matrices. As is proven in Ref.\cite{CM92b}, this model is equivalent as
$N\ra\infty$ to the following model in an external field:
\be
Z_N [\h;\alpha] = \e^{-\fr N2 \tr{\h^2}} \int DX\,
\e^{N\tr{\left(-\fr 12 X^2+\h X+\alpha\log X\right)}},
\label{2}
\ee
with $\h$ and $\{t_k\}$ being related by the Miwa transformation
\be
t_k= \fr{1}{k} \tr{\h^{-k}}-\fr N2 \delta_{k2} \hbox{\ \ \ \ for \ \ }
k\geq1,\,\,\,\,\,\,\, t_0=\hbox{\rm tr}\,\log{\h^{-1}} .
\label{Miwa}
\ee
The partition function (\ref{2}), in turn, is associated \cite{CM92a} with
an external field problem for the Penner model \cite{Pen86}:
\be
Z_N [\L;\alpha] = \e^{-\fr {\alpha^2N}{2}\tr{\L^{-2}}}
(\det{\L})^{N(\alpha+1)}    \int DX\,
\e^{N\tr{\left(-\fr 12 \L X\L X+\alpha[\log{(1+X)}-X]\right)}}
\label{3}
\ee
providing $\h=\L-\alpha\L^{-1}$. The extra coefficients are introduced
to provide
\be
Z_{\alpha N} [t_.]= Z_N [\h;\alpha]=Z_N [\L;\alpha]
\ee
to any order of the genus expansion.

A surprising property of \hm, which is advocated in
\mbox{Refs.\cite{CM92b,CM92a}},
is that it reveals some features of $c=1$ CFT interacting with 2D
gravity. For the case of the Penner model, this property was discovered
by Distler and Vafa \cite{DV91} in the {\it double-scaling limit\/}
and has been
extended to more general models by Tan \cite{Tan91a},
by Chaudhuri, Dykstra and Lykken \cite{CDL91}
and by Gilbert and Perry \cite{GP91}, in particular to the model
(\ref{3}) with $\L$ being proportional to a unit matrix. However,
these features of $c=1$ have been observed in Refs.\cite{CM92b,CM92a}
in genus zero and genus one, identifying $\alpha$ with the cosmological
constant, {\it without\/} taking the double scaling limit. Therefore, it
was conjectured that \hm\ has something to do with the {\it continuum\/}
$c=1$ case similar to the Kontsevich model \cite{Kon91} which is
associated with the continuum pure 2D gravity ($c=0$) or to its
generalizations \cite{KMMMZ91a} which are associated with $c<1$.

A direct way to verify this conjecture is to study correlation functions
of the loop operators. For \hm\ with an arbitrary (not necessarily
symmetric) potential, the one-loop correlator has been known in genus zero for
a long time \cite{Mig83}, while the two- and three-loop correlators
were obtained in Ref.\cite{AJM90}. The genus one contribution to the
one-loop correlator was explicitly calculated in Ref.\cite{AM90} for a
quartic symmetric potential and in Ref.\cite{CM92b} for an arbitrary
symmetric potential. On the other hand, much is known now about $c=1$
correlators since pioneering works by Kostov \cite{Kos88} and Boulatov
\cite{Bou90}. An incomplete list of references includes
\cite{Par90}--\cite{MSW91}.

In the present paper
we develop an iterative algorithm for calculating the genus expansion
of the hermitian $N\times N$ one-matrix model ($=$ the Penner model in
an external field). We introduce the moments, $I_p$ and
$J_p$, of the external field
as well as the `basis vectors', $\chi^{(n)}$ and $\psi^{(n)}$, which
are determined by a recursion relation and diagonalize the iterative
procedure. We prove that
the genus $g$ contribution to the partition function depends only on
$I_p$ and $J_p$ with $p\leq 3g-2$ while the genus $g$ contribution
to the $m$-loop correlator depends on $I_p$ and $J_p$ with $p\leq
3g-2+m$.
We present the explicit calculation of the partition function and the
one-loop correlator for an arbitrary potential in genus one.
We compare the correlators
for the hermitian one-matrix model with those at zero momenta for $c=1$ CFT
and show an agreement of the one-loop correlators
(this correlator in $c=1$ is the only one which vanishes at non-zero momenta)
in genus zero.
We did not find such an agreement for the two-loop correlator.

\newsection{The iterative scheme}

We propose in this section a general iterative procedure for calculating
higher genus contributions to \hm. We shall be solving, iteratively
in $1/N^2$, the loop equation
\be
\int_{C_1} {d\omega\over 2\pi i}{V^\prime(\omega)\over (\l-\omega)}
W(\omega)=(W(\l))^2 + W(\l;\l)\, ,
\label{le}
\ee
where
\be
V(\l)= \sum_{k=0}^\infty t_k \l^k,\,\,\,\,\,
{\delta \over \delta V(\lambda)}= - \sum_{k=0}^\infty \lambda^{-k-1} \ddt{k}
\label{ddV}
\ee
and
\be
W(\l)= {\delta \over \delta V(\lambda)} \log{Z[t_.]},\,\,\,\,\,
W(\l;\l)= {\delta \over \delta V(\lambda)} W(\l)
\label{W}
\ee
with the partition function $Z[t_.]$ given by Eq.(\ref{1}).
The one-loop correlator $W(\l)$ obeys the normalization condition
\be
\l W(\l)\ra \alpha N \hbox{\ \ \ \ \ as \ \ }\l\ra\infty .
\label{bound}
\ee
While we consider in this paper the loop equation (\ref{le}), it should
be noticed that a similar iterative
procedure can be formulated for the equivalent Schwinger--Dyson equation
\be
\left\lbrace {\d^2 \over \d\h_i^2}+\sum_{j\neq i} \frac {1}{\h_i-\h_j}
\left(\dd{\h_i}-\dd{\h_j}\right)
+ N\h_i\dd{\h_i} -\alpha N^2 \right\rbrace
Z[\h_.;\alpha] = 0
\label{sd}
\ee
where the partition function $Z[\h_.;\alpha]$ is given by Eq.(\ref{2})
and $\{t_k\}$ are related to $\h_i$ --- the eigenvalues of $\h$ --- by
Eq.(\ref{Miwa}).
The genus one solution of Ref.\cite{CM92b} was obtained for a symmetric
distribution of the eigenvalues by solving Eq.(\ref{sd}).

Our goal is to solve Eq.(\ref{le}) iteratively order by order in
$1/N^2$ starting with the one-cut genus zero solution \cite{Mig83}
\be
W_0(\l)=\int_{C_1}{d\omega\over 4\pi i}{V^\prime(\omega)\over (\l-\omega)}
\frac {\qqqq \l}{\qqqq \omega},
\label{W0}
\ee
with $b$ and $c$ given by
\be
\int_{C_1}{d\omega\over 2\pi i}{V^\prime(\omega)\over \qqqq \omega } =0,
\,\,\,\,\,\,
\int_{C_1}{d\omega\over 2\pi i}{\omega V^\prime(\omega)\over \qqqq \omega }
=2 \alpha N .
\label{bandc}
\ee
In this way we will obtain the genus expansion of the (logarithm of the)
partition function
\be
\log Z[t_.] = \sum_{g=0}^\infty  F_g[t_.]
\label{F}
\ee
and that of the associated multi-loop correlators.

Following Ref.\cite{CM92b}, we introduce the new variables
\bea
I_p&=& \int_{C_1}{d\omega\over 2\pi i}{\omega V^\prime(\omega)
\over \left(\omega^2+b\omega+c\right)^{p+\half}}
-(2\alpha + 1)N \delta_{p0} , \nonumber \\
J_p&=& \int_{C_1}{d\omega\over 2\pi i}{V^\prime(\omega)\over
\left(\omega^2+b\omega+c\right)^{p+\half}}.
\label{nv}
\eea
Their relation to the moments of the external field $\h$, which is
described by the density of eigenvalues $\rho(x)$,
can easily be established by substituting
\be
\frac 1N V^\prime (\l)=
\pint dx {\rho(x)\over x-\l} -\l
\label{hilbert}
\ee
into Eq.(\ref{nv}):
\bea
\frac 1N I_p& =& \int dx {x\rho(x)\over {\ppp x}^{p+\half}}
-(\fr{3}{8}b^2-\fr{1}{2}c)\delta_{p0}-\delta_{p1},
\nonumber \\
\frac 1N J_p&=& \int dx {\rho(x)\over {\ppp x}^{p+\half}}
 -\fr{1}{2}b\delta_{p0}.
\label{moments}
\eea
The terms with $\delta_{p0}$ and
$\delta_{p1}$ come from the second term on the r.h.s.\ of
Eq.(\ref{hilbert}) taking the residual at infinity.
It is easy to see, expanding the integrands in Eq.(\ref{moments}) in
$1/x$ that $I_p$ depends on $t_k$ with $k\geq 2p$ while $J_p$
depends on $t_k$ with $k\geq 2p+1$.

As has been proposed in Ref.\cite{CM92b}, $F_g$ depends
at $1\leq g <\infty$ only on $I_p$ and $J_p$ for $p\leq P_g-1$ (where
$P_g=3g-1$ as is proven below). This is in contrast
to the $t$-dependence of $F_g$ which always depends on the whole set
$\{t_k\}$.

For an iterative solution of the loop equation, it is convenient to
introduce the operator $\hat{\KK}$ by%
\footnote{Inserting here Eq.(\ref{hilbert}), one finds the relation with the
operator $\KK$ from Ref.\cite{CM92b}: $-\frac 1N \hat{\KK}= \KK+\l$,
when applied to a function $\Phi(\l)$ which decays at infinity faster
than $1/\l$ (if $\Phi(\l)\sim 1/\l$ as $\l\ra\infty$, $-1$ should be
added to the r.h.s.\ which comes from the residual at infinity).}
\be
\hat{\KK}\Phi(\l)= \int_{C_1} {d\omega\over 2\pi i}
{V^\prime(\omega)\over(\l -\omega)}\Phi(\omega).
\label{K}
\ee

It is easy to calculate of how the operator $\hat{\KK}$ acts on the set
of functions
\be
\Phi^{(n)}(\l)=\frac{1}{{\ppp \l}^{n+\half}}
\ee
as well as on $\l\Phi^{(n)}(\l)$. Let us start with $n=0$. Comparing with
Eq.(\ref{W0}), one gets immediately
\be
\hat{\KK}\Phi^{(0)}(\l)= 2 W_0(\l)\Phi^{(0)}(\l).
\ee
To calculate of how $\hat{\KK}$ acts on $\Phi^{(n)}(\l)$ with $n\geq 1$,
let us insert the following expansion of the denominator
\be
{1\over \l-\omega}=
(\l+\omega+b)\sum_{k=1}^{n}{{\ppp \omega}^{k-1}\over{\ppp \l}^{k}}
+{{\ppp \omega}^n\over{\ppp \l}^n}{1\over \l-\omega} \hbox{\ \ for \ }
n\geq 1.
\label{prog}
\ee
Using Eq.(\ref{nv}) we get finally
\be
\hat{\KK}\Phi^{(n)}(\l)= \int_{C_1} {d\omega\over 2\pi i}{
V^\prime(\omega)\over(\l-\omega)}\Phi^{(n)}(\omega) =
\sum_{k=1}^{n}\frac{I_{n+1-k}+(\l+b)J_{n+1-k}}{{\ppp \l}^{k}}
+ 2 W_0(\l)\Phi^{(n)}(\l).
\label{finally}
\ee
Quite similarly, one calculates
of how $\hat{\KK}$ acts on $\l\Phi^{(n)}(\l)$:
\bea
& &\hat{\KK}\l\Phi^{(n)}(\l)=-J_n+\l\hat{\KK}\Phi^{(n)}(\l)= \nonumber \\
&=&\sum_{k=1}^{n}\frac{\l I_{n+1-k}-cJ_{n+1-k}}
{{\ppp \l}^{k}}+\sum_{k=1}^{n-1}\frac{J_{n-k}}{{\ppp \l}^{k}}
+ 2 W_0(\l)\l\Phi^{(n)}(\l).
\label{finally'}
\eea

Some comments concerning these formulas are in order. The expansion is
in integer powers of $1/\ppp \l$ due to Eq.(\ref{prog}). The sum over
$k$ is bounded from above by $n$ by construction.
For the symmetric potential $V(\l)=V(-\l)$, which
corresponds to a symmetric $\rho(\l)=\rho(-l)$ in Eq.(\ref{hilbert}),
when $b=0$ and $J_p=0$, Eq.(\ref{finally}) recovers Eq.(5.13) of
Ref.\cite{CM92b}.

Let us now turn to the iterative solution of Eq.(\ref{le}). Substituting
the genus expansion for $W(\l)$ and $W(\l;\l)$:
\be
W(\l)=\sum_{g=0}^\infty W_g(\l)\,;\,\,\,\,
W(\l;\l)=\sum_{g=0}^\infty W_g(\l;\l),
\ee
and comparing the terms of the same order in $1/N^2$, one rewrites
Eq.(\ref{le}) as
\be
\left(\hat{\KK}- 2 W_0(\l)\right)W_g(\l)=
\sum_{g^{\p}=1}^{g-1} W_{g^{\p}}(\l)W_{g-g^{\p}}(\l) +W_{g-1}(\l;\l)
\label{leit}
\ee
where $ g\geq 1$. This equation allows us to calculate $W(\l)$ order by
order of genus expansion.

As is proven below by induction, the r.h.s.\  of Eq.(\ref{leit}),
which is known from previous orders of genus expansion, has a structure
of the sum over $n$ up to
\be
n_g=3g-1
\ee
of the terms of the type
$A_g^{(n)}/{\ppp \l}^n$ and $D_g^{(n)}\l/{\ppp \l}^n$ with the coefficients
$A_g^{(n)}$ and $D_g^{(n)}$ being some functions of the moments $I_p$ and
$J_p$ to be discussed latter. For this reason, it is convenient to introduce
two sets of `basis vectors' $\chi^{(n)}(\l)$ and $\psi^{(n)}(\l)$ which
obey
\be
\left(\hat{\KK}- 2 W_0(\l)\right)\chi^{(n)}(\l)= {1\over{\ppp \l}^n},\,\,\,
\left(\hat{\KK}- 2 W_0(\l)\right)\psi^{(n)}(\l)= {\l\over{\ppp \l}^n}.
\label{vectors}
\ee
This idea is
analogous to the one suggested by Gross and Newman \cite{GN91}
for the unitary matrix model and for the hermitian model with
a cubic potential.

We can derive now recursion relations
which defines  $\chi^{(n)}(\l)$ and $\psi^{(n)}(\l)$ explicitly. Using
Eqs.(\ref{finally}) and (\ref{finally'}), we get
\bea
\chi^{(n)}(\l)&=&\frac {I_1-\l J_1}{\Delta}\Phi^{(n)}(\l)-
\nonumber \\ & &-\frac {1}{\Delta}\sum_{k=1}^{n-1}
\chi^{(k)}(\l)\{I_1[I_{n+1-k}+bJ_{n+1-k}]+J_1[cJ_{n+1-k}-J_{n-k}]\}
- \nonumber \\
& &- \frac {1}{\Delta}\sum_{k=1}^{n-1}\psi^{(k)}(\l)\{I_1J_{n+1-k}-J_1I_{n+1-k}
  \}
\label{chi}
\eea
and
\bea
\psi^{(n)}(\l)&=& \frac {\l(I_1+bJ_1)+cJ_1}{\Delta}\Phi^{(n)}(\l)-\nonumber \\
 & &- \frac {1}{\Delta}\sum_{k=1}^{n-1}\chi^{(k)}(\l)
\{cJ_1[I_{n+1-k}+bJ_{n+1-k}]+(I_1+bJ_1)[J_{n-k}-cJ_{n+1-k}]\}
-\nonumber \\ & &- \frac {1}{\Delta}\sum_{k=1}^{n-1}\psi^{(k)}(\l)
\{cJ_1J_{n+1-k}+(I_1+bJ_1)I_{n+1-k}  \}
\label{psi}
\eea
where
\be
\Delta = I_1^2 + bI_1J_1 + cJ_1^2 .
\label{Delta}
\ee

Eqs.(\ref{vectors}) and (\ref{chi}), (\ref{psi}) allow us to restore $W_g(\l)$
provided the r.h.s.\ of Eq.(\ref{leit}) has the advertised form.
The result reads
\be
W_g(\l)=\sum_{n=1}^{n_g} \left[ A_g^{(n)} \chi^{(n)}(\l)
+  D_g^{(n)} \l\psi^{(n)}(\l) \right].
\label{ansatz}
\ee
Notice that we do not add the terms with $n=0$ which are annihilated by
the operator $\hat{\KK}-2W_0(\l)$. These terms would contradict the
boundary condition (\ref{bound}). The expression on the r.h.s.\ of
Eq.(\ref{ansatz}) is analytic everywhere in the complex $\l$-plane except for
the cut coinciding with the one of $W_0(\l)$. This is in accordance with the
different iterative
procedure advocated by Migdal \cite{Mig83} and elaborated by David
\cite{Dav90}.

It remains to prove that the r.h.s.\ of Eq.(\ref{leit}) indeed has such
a form. We see from Eq.(\ref{ansatz}) that the first term on the r.h.s.\
of Eq.(\ref{leit}) has exactly this form order by order in the genus
expansion, while the second term needs a more careful analysis since we
have to calculate $\delta W_{g-1}(\l)/\delta V(\l)$. To do this we need
some more formulas.

 From Eq.(\ref{nv}) one gets
\bea
\frac {\d I_p}{\d b}= - (p+\fr 12)[bI_{p+1}+cJ_{p+1}+J_{p}],\,\,\,
\frac {\d I_p}{\d c}= - (p+\fr 12)I_{p+1} ; \nonumber \\
\frac {\d J_p}{\d b}= - (p+\fr 12)I_{p+1},\,\,\,
\frac {\d J_p}{\d c}= - (p+\fr 12)J_{p+1}
\label{dIJdbc}
\eea
while Eq.(\ref{bandc}), which determines $b$ and $c$, can be rewritten as
\be
I_0=-N,\,\,\,\,\,J_0=0.
\label{bandc'}
\ee
One needs as well the following formulas for calculating
$\delta /\delta V(\l)$:
\bea
{\delta b\over\delta V(\l)}& =& \dd \l \frac {2}{\qqqq \l}
\frac {I_1-\l J_1}{\Delta}, \nonumber \\
{\delta c\over\delta V(\l)}&=& \dd \l \frac {2}{\qqqq \l}
\frac {(\l+b)I_1+cJ_1}{\Delta},
\label{dbcdV}
\eea
with $\Delta$ given by Eq.(\ref{Delta}), and
\bea
{\delta I_p\over\delta V(\l)}&=& \dd \l \l\Phi^{(p)}(\l)
-{\delta b\over\delta V(\l)} (p+\fr 12)[bI_{p+1}+cJ_{p+1}+J_{p}]
-{\delta c\over\delta V(\l)} (p+\fr 12)I_{p+1}, \nonumber \\
{\delta J_p\over\delta V(\l)}&=& \dd \l  \Phi^{(p)}(\l)
-{\delta b\over\delta V(\l)} (p+\fr 12)I_{p+1}
-{\delta c\over\delta V(\l)} (p+\fr 12)J_{p+1}.
\label{dIJdV}
\eea
These formulas%
\footnote{We have not calculated explicitly $\d/\d \l$
in order that $\delta /\delta \rho(\l)$, which is
related to $\delta /\delta V(\l)$ by
$$
{\delta \over\delta V(\l)}=\frac 1N \dd \l {\delta \over\delta \rho(\l)},
$$
could be extracted easily as well.}
can be obtained from Eqs.(\ref{nv}) and (\ref{bandc'})
using Eq.(\ref{dIJdbc}).

It is easy to see that the result of acting of $\delta /\delta V(\l)$
on $W_{g-1}$ given by Eq.(\ref{ansatz}) has exactly the form discussed
above. Moreover, we have presented an explicit algorithm for
calculating $W_g(\l)$, say by symbolic computer calculations with the
only input parameter being $g$. The explicit results for genus one are
presented in the next section.

We perform now a power-counting analysis to express $n_g$ and $P_g$ ,
which are introduced above, via $g$. Let us note first that the highest
moments $I_{P_g}$ and $J_{P_g}$ emerge on the r.h.s.\ of Eq.(\ref{ansatz})
from the highest term $1/{\ppp \l}^{n_g}$ on the r.h.s.\ of
Eq.(\ref{leit}) according to Eqs.(\ref{chi}), (\ref{psi}) (they are
associated with the $k=1$ terms). Therefore, one gets $P_g=n_g$. To pass
to the next order of the genus expansion, one analyze the structure of
the r.h.s.\ of Eq.(\ref{leit}). It is easy to estimate the highest
power of each term $W_{g^{\p}}(\l)W_{g-g^{\p}}(\l)$ for the solution
(\ref{ansatz}), which is known from the previous order, to be
$n_{g^{\p}}+n_{g-g^{\p}}+1$ while the two-loop correlator gives the
power $P_g+3$ by virtue of Eqs.(\ref{dbcdV}) and (\ref{dIJdV}).
Therefore, one gets $n_{g+1}=n_g+3$ and finally
\be
P_g=3g-1
\ee
since $n_1=2$.
Notice that both types of terms contribute to the maximal power
since $n_{g^{\p}}+n_{g-g^{\p}}+1=3g-1$.

According to Eqs.(\ref{F}), (\ref{W}) and (\ref{dbcdV}), (\ref{dIJdV}),
the highest moments which contribute to $F_g$ are $I_{3g-2}$ and
$J_{3g-2}$. An analogous result for the Kontsevich model has been
obtained by Itzykson and Zuber \cite{IZ92}.%
\footnote{It would be suggestive to relate these numbers to the
(complex) dimension of the moduli space ${\cal M}_{g,m}$, which
equals $3g-3+m$.}
Applying Eqs.(\ref{dbcdV}),
(\ref{dIJdV}) $m$ times to obtain the $m$-loop correlator, one finds
highest moments to be $I_{3g-2+m}$ and $J_{3g-2+m}$.
Thus, we have proven the following
\begin{quotation}
{\bf Theorem} \ \  {\it The genus $g$ contribution to the partition function
depends on $I_p$ and $J_p$ with $p\leq 3g-2$. The genus $g$ contribution
to the $m$-loop correlator depends on $I_p$ and $J_p$ with $p\leq
3g-2+m$}.
\end{quotation}

\newsection{The partition function and correlators in genus one}

We present in this section the explicit results for the genus one
one-loop correlator $W_1(\l)$  and partition function $F_1$
for the case of an arbitrary potential $V(\l)$ which are
obtained according to the algorithm of the previous section.

In genus one there is no sum on the r.h.s.\ of Eq.(\ref{leit}) while
\cite{AJM90}
\be
W_0(\l;\l)=\frac {b^2-4c}{16{\ppp \l}^2},\,\,\,\,
W_0(\l;\mu)= \frac{1}{2(\l-\mu)^2}\left[
\frac{\l\mu-\hf b(\l+\mu)+c}{\qqqq \l\,\,\qqqq \mu}-1\right].
\label{two-loop}
\ee
Eq.(\ref{vectors}) yields immediately
\bea
W_1(\l)=\frac {b^2-4c}{16}\chi^{(2)}(\l)
= \frac {b^2-4c}{16}\left\lbrace \frac{\Phi^{(2)}(\l)}{\Delta}
[I_1-\l J_1] - \right.\nonumber \\ -\frac{\Phi^{(1)}(\l)}{\Delta^2}
[-J_1^2I_1+I_1^2I_2+bI_1^2J_2-cJ_1^2I_2+2cI_1J_1J_2]- \nonumber \\
\left.-\frac{\l\Phi^{(1)}(\l)}{\Delta^2}
[J_1^3+I_1^2J_2-2I_1J_1I_2-bJ_1^2I_2-cJ_1^2J_2]
\right\rbrace
\label{W1}
\eea
where we have substituted the explicit form of $\chi^{(2)}(\l)$
given by Eqs(\ref{chi}) and (\ref{psi}). For the symmetric case when
$b=0$ and $J_p=0$, Eq.(\ref{W1}) recovers Eq.(5.18) of Ref.\cite{CM92b}.

We are now in a position to integrate Eq.(\ref{W1}) (w.r.t.\ $\rho$) and
find the functional $F_1$ whose variation generates this $W_1(\l)$.
First, we integrate over $\l$ using the formulas \pagebreak
\bea
\fr 14 \int {dx\over {\ppp x}^{3/2}}&=& -{x+\hf b\over (b^2-4c)\qqqq x}
+\hbox{\ const},
\nonumber\\
\fr 14 \int {dx\over {\ppp x}^{5/2}}&=& -{x+\hf b\over 3(b^2-4c){\ppp x}^{3/2}}
+ \nonumber\\
& &+ {8(x+\hf b)\over 3(b^2-4c)^2\qqqq x}+\hbox{\ const}.
\label{A10}
\eea
It is worth to note the appearance of the factor $b^2-4c$ in the denominators.

Using Eqs.(\ref{dIJdV}) and (\ref{dbcdV}),
we can now integrate $W_1(\l)$
explicitly. Namely, it is more or less clear that one can produce
coefficients like in Eq.(\ref{W1}) by differentiating terms like $\log \Delta$
(the terms with $\Delta^2$ in the denominators  originate from the
variations (\ref{dIJdV}), (\ref{dbcdV})\/). Moreover, due to the appearance of
${1\over\Delta (b^2-4c)}$ term after the integration of $W_1(\l)$ over $\l$,
it is reasonable to assume the presence of $\log (4c-b^2)$ term in $F_1$.
After all these preliminaries, a straightforward, though
lengthy, calculation gives the answer
\be
F_1 = -\frac{1}{12}\log (4c-b^2) -\frac{1}{24} \log \Delta + \hbox{\
const}\,,
\label{A13}
\ee
which, in particular,  reproduces Eq.(5.18) of Ref.\cite{CM92b} for the reduced
model (the symmetric potential).

\newsection{Comparison with $c=1$ correlators}

We compare in this section the explicit formulas for the loop
correlators in \hm\ with those in $c=1$ CFT. A problem immediately
arises that the loop operator in $c=1$ depends both on the length of the
loop $l$ and on momentum $p$ which is associated with a matter field.
Since in \hm\ there is no dependence on $p$, we should put $p$ equal to
some value (or sum up somehow over special values of $p$). The procedure
is, however, unambiguous for the one-loop correlator which is
non-vanishing only at $p=0$ due to the momentum conservation. Therefore,
our idea is to compare first the one-loop correlators.

The genus zero one-loop correlator, given by Eqs.(\ref{W0}) and
(\ref{bandc}), is greatly simplified after differentiating w.r.t.\ the
cosmological constant $\alpha$:
\be
\frac 1N \frac {d W_0(\l)}{d \alpha}= \frac{1}{\qqqq \l}.
\ee
Let us put for simplicity $b=0$ keeping in mind that the general case
can be restored by the shift $\l\ra\l+b/2$; $c\ra c-b^2/4$.
The equation (\ref{bandc}) which determines $c$ can be written in the
form of the genus zero string equation
\be
\sum_m mt_{2m}\frac {(2m-1)!!}{(2m)!!}(-c)^m = \alpha.
\label{se}
\ee
If we restrict ourselves to the simplest case $t_2 \neq 0$, $t_{2m}=0$
for $m\geq 2$, Eq.(\ref{se}) gives $c=-\alpha/t_2$ and $c>0$ corresponds
to the `upside-down' potential which is familiar from the quantum
mechanical description of c=1 correlators. Moreover, the r.h.s.\ of
Eq.(\ref{se}) is as well the known expression for the (diagonal)
resolvent of the Schr\"{o}dinger operator with the potential $c(\alpha)$
in genus zero. Therefore, we conclude that the genus zero one-loop
correlators coincide.

The genus zero two-loop correlator in c=1 theory at zero momentum can easily
be extracted, applying the Gegenbauer's addition theorem,
from the results by Moore and Seiberg \cite{Moo92}%
\footnote{We thank G.Moore for a e-mail correspondence concerning this
formula}:
\be
W_0(l_1,p=0;l_2,p=0) = K_0\left(\sqrt{c}(l_1+l_2)\right)
\label{KO}
\ee
where $K_0$ is the modified Bessel function of the third kind.

It is easy to see that the Laplace transform of Eq.(\ref{KO}) does not
coincide with Eq.(\ref{two-loop}). A possible way out (a pure
speculative one) might be to sum up the general formula of Moore and
Seiberg \cite{Moo92} over the momenta which corresponds to the discrete
special states of $c=1$ CFT \cite{GKN91,Wit91c,KP91}. This problem
deserves further investigation.

\section*{Acknowledgments}

We thank C.Kristjansen for discussions. Yu.M. thanks the NBI high energy
theory group for the hospitality at Copenhagen.


\begin{thebibliography}{10}
\small
\addtolength{\itemsep}{-6pt}

\bibitem{CM92b}
L.Chekhov and Yu.Makeenko, {\it A hint on the external field problem for matrix
  models}, {\sl preprint~NBI-HE-92-06} (January, 1992), {\sl Phys.Lett.}
  {\bf B} in print.

\bibitem{CM92a}
L.Chekhov and Yu.Makeenko, {\it The multicritical Kontsevich--Penner model},
  {\sl preprint~NBI-HE-92-03} (January, 1992).

\bibitem{Pen86}
R.C.Penner, {\sl Bull.Am.Math.Soc.}, {\bf 15} (1986) 73;
  {\sl J.Diff.Geom.}, {\bf 27} (1988) 35.

\bibitem{DV91}
J.Distler and C.Vafa, {\sl Mod.Phys.Lett.}, {\bf A6} (1991) 259.

\bibitem{Tan91a}
C.-I~Tan, {\sl Mod.Phys.Lett.}, {\bf A6} (1991) 1373;
  {\it Generalized Penner models and multi-critical behavior}, {\sl
  preprint~BROWN-HET-810} (May, 1991).

\bibitem{CDL91}
S.Chaudhuri, H.Dykstra, and J.Lykken, {\sl Mod.Phys.Lett.}, {\bf A6} (1991)
  1665.

\bibitem{GP91}
G.Gilbert and M.Perry, {\sl Nucl.Phys.}, {\bf B364} (1991) 734.

\bibitem{Kon91}
M.L.Kontsevich, {\sl Funk.Anal.\&Prilozh.}, {\bf 25} (1991) 50 (in Russian).

\bibitem{KMMMZ91a}
S.Kharchev, A.Marshakov, A.Mironov, A.Morozov, and A.Zabrodin, {\it Unification
  of all string models with $c<1$}; {\it Towards unified theory of 2d gravity},
  {\sl preprints~FIAN/ITEP} (October, 1991).

\bibitem{Mig83}
A.A.Migdal, {\sl Phys.Rep.}, {\bf 102} (1983) 199.

\bibitem{AJM90}
J.Ambj{\o}rn, J.Jurkiewicz, and Yu.Makeenko, {\sl Phys.Lett.}, {\bf 251B}
  (1990) 517.

\bibitem{AM90}
J.Ambj{\o}rn and Yu.Makeenko, {\sl Mod.Phys.Lett.}, {\bf A5} (1990) 1753.

\bibitem{Kos88}
I.K.Kostov, {\sl Phys.Lett.}, {\bf 215B} (1988) 499.

\bibitem{Bou90}
D.V.Boulatov, {\sl Phys.Lett.}, {\bf 237B} (1990) 202.

\bibitem{Par90}
G.Parisi, {\sl Phys.Lett.}, {\bf 238B} (1990) 209.

\bibitem{GK90}
D.Gross and I.Klebanov, {\sl Nucl.Phys.}, {\bf B344} (1990) 475.

\bibitem{B-M91}
S.Ben-Menahem, {\sl Nucl.Phys.}, {\bf B364} (1991) 681.

\bibitem{GKN91}
D.Gross, I.Klebanov, and M.Newman, {\sl Nucl.Phys.}, {\bf B350} (1991) 621; \\
U.Danielsson and D.Gross, {\sl Nucl.Phys.}, {\bf B366} (1991) 27; \\
U.Danielsson, {\it Symmetries and special states in two dimensional string
  theory}, {\sl preprint~PUPT-1301} (December, 1991).

\bibitem{KL91}
I.Klebanov and D.Lowe, {\sl Nucl.Phys.}, {\bf B363} (1991) 543.

\bibitem{DJ90}
S.Das and A.Jevicki, {\sl Mod.Phys.Lett.}, {\bf A5} (1990) 1639.

\bibitem{DJR91a}
K.Demeterfi, A.Jevicki, and J.Rodrigues, {\sl Nucl.Phys.}, {\bf B362} (1991)
  173; {\sl Nucl.Phys.}, {\bf B362} (1991)
  499; {\sl Mod.Phys.Lett.}, {\bf A6} (1991) 3199.

\bibitem{Moo92}
G.Moore, {\sl Nucl.Phys.}, {\bf B368} (1992) 557; \\
G.Moore, N.Seiberg, and M.Staudacher, {\sl Nucl.Phys.}, {\bf B362} (1991)
  665; \\
E.Martinec, G.Moore, and N.Seiberg, {\sl Phys.Lett.}, {\bf 263B} (1991)
  190; \\
G.Moore and N.Seiberg, {\it From loops to fields in 2d quantum gravity}, {\sl
  preprint~RU-91-29, YCTP-P19-91} (July, 1991).

\bibitem{Kos91a}
I.K.Kostov, {\sl Phys.Lett.}, {\bf 266B} (1991) 42;
  {\sl Phys.Lett.}, {\bf 266B} (1991) 317;
  {\it Strings with discrete target space}, {\sl
  preprint~SPhT/91-142} (September, 1991).

\bibitem{MSW91}
G.Mandal, A.Sengupta, and S.Wadia, {\sl Mod.Phys.Lett.}, {\bf A6} (1991)
  1465; \\
S.Das, A.Dhar, G.Mandal, and S.Wadia, {\it Bosonization of nonrelativistic
  fermions and W-infinity algebra}, {\sl preprint~IASSNS-HEP-91/72} (November,
  1991);
  {\it W-infinity ward identities and
  correlation functions in the $c=1$ matrix model}, {\sl
  preprint~IASSNS-HEP-91/79} (December, 1991).

\bibitem{GN91}
D.J.Gross and M.J.Newman, {\sl Phys.Lett.}, {\bf 266B} (1991) 291.

\bibitem{Dav90}
F.David, {\sl Mod.Phys.Lett.}, {\bf A5} (1990) 1019.

\bibitem{IZ92}
C.Itzykson and J.-B.Zuber, {\it Combinatorics of the modular group II. The
  Kontsevich integrals}, {\sl preprint~SPhT/92-001} (January, 1992).

\bibitem{Wit91c}
E.Witten, {\it Ground ring of two dimensional string theory}, {\sl
  preprint~IASSNS-HEP-91/51} (August, 1991).

\bibitem{KP91}
I.R.Klebanov and A.M.Polyakov, {\sl Mod.Phys.Lett.}, {\bf A6} (1991) 3273; \\
I.R.Klebanov, {\it Ward identities in two-dimensional string theory}, {\sl
  preprint~PUPT-1302} (December, 1991).

\end{thebibliography}
\end{document}